\documentstyle[prl,twocolumn,aps,epsf]{revtex}
\begin{document}
\draft

\twocolumn[\hsize\textwidth\columnwidth\hsize\csname
@twocolumnfalse\endcsname

\title{\bf  Evidence for two  superconducting energy gaps  in MgB$_2$ by
point-contact spectroscopy}
\author{P.  Szab\'o,$^{1,2}$  P.  Samuely,$^{1}$  J.  Ka\v cmar\v
c\'{\i}k,$^{1}$ T. Klein,$^{2}$, J. Marcus,$^{2}$ D. Fruchart, $^{3}$
S. Miraglia,$^{3}$ C. Marcenat,$^{4}$ A. G. M. Jansen$^{5}$}
\address{$^1$Institute of Experimental Physics, Slovak Academy of
Sciences,         SK-04353         Ko\v{s}ice,         Slovakia\\
$^2$Laboratoire d'Etudes des Propri\'et\'es Electroniques des Solides,
Centre  National  de  la  Recherche  Scientifique, B.P. 166,
F-38042 Grenoble Cedex 9, France\\
$^3$Laboratoire de  Cristallographie, Centre National  de la
Recherche Scientifique, BP 166, 38042 Grenoble Cedex 9, France\\
$^4$CEA-Grenoble, D\'epartement de Recherche Fondamental sur
la Mati\'ere Condens\'ee, F-38054 Grenoble Cedex 9, France\\
$^5$Grenoble       High       Magnetic      Field      Laboratory
Max-Planck-Institut     f\"{u}r     Festk\"{o}rperforschung    and
Centre National  de la Recherche Scientifique,  B.P. 166, F-38042 Grenoble Cedex 9, France}

\date{\today}
\maketitle

\begin{abstract}
Experimental support  is found for  the multiband model  of
the  superconductivity  in  the  recently  discovered system
MgB$_2$  with the  transition temperature  $T_c$ =  39 K. By
means  of Andreev  reflection evidence  is obtained  for two
distinct superconducting  energy gaps. The sizes  of the two
gaps  ($\Delta_S$ =  2.8 meV   and $\Delta_L$  = 7  meV) are
respectively  smaller  and  larger  than  the  expected weak
coupling  value.  Due  to  the  temperature  smearing of the
spectra the two gaps  are hardly distinguishable at elevated
temperatures  but  when  a  magnetic  field  is  applied the
presence of two  gaps can be demonstrated close  to the bulk
$T_c$ in the raw data.

\end{abstract}

]



Two decades  of the boom  in the field  of superconductivity
has  recently been  boosted by  the surprising  discovery of
superconductivity in MgB$_2$ \cite{akimitsu}. In contrast to
the        cuprates,        the        first       tunneling
\cite{rubio,karapetrov,zasadzinsky}     and    point-contact
\cite{kohen,plecenik}    spectroscopy    measurements   have
unequivocally   shown  that   this  system   is  a  $s$-wave
superconductor  and isotope  effects \cite{budko,hinks} have
pointed towards a phonon mechanism. However, the size of the
superconducting energy  gap has remained  unclear. We report
here  on experimental  support  for  the multiband  model of
superconductivity   recently   proposed   by   Liu  {\it  et
al.}\cite{liu}  thus  showing  that  MgB$_2$  belongs  to an
original class  of superconductors in which  two distinct 2D
and  3D  Fermi  surfaces  contribute  to  superconductivity.
Indeed,  our point-contact  spectroscopy experiments clearly
show the existence of two distinct superconducting gaps with
$\Delta_S$(0) = 2.8 meV and $\Delta_L$(0) = 7 meV. Both gaps
close near to the bulk  transition temperature $T_c$ = 39 K.
Our measurements in magnetic field  show directly in the raw
data  the  presence  of  two  superconducting  gaps  at  all
temperatures up to the same bulk transition $T_c$ indicating
that the  two gaps are inherent  to the superconductivity in
MgB$_2$.

Although    quite   scattered,    the   first   spectroscopy
measurements \cite{rubio,karapetrov,zasadzinsky,kohen,plecenik,takahashi}
yielded to  superconducting gap  values
surprisingly  smaller  than  the  BCS  weak  coupling  limit
$2\Delta/kT_c$  =  3.52.  Moreover  simultaneous topographic
imaging   and  quasiparticle   density  of   states  mapping
\cite{giubileo} revealed substantial  inhomogeneities at the
surface of the  sample as well as a  large scattering of the
energy  gap  values  measured  at  different  parts  of  the
polycrystalline sample (with $\Delta$  ranging from 3 to 7.5
meV). This  energy gap distribution can  be caused by sample
inhomogeneities.  However, Giubileo  {et al.}  also observed
a superposition  of two  gaps ($\Delta_S$(0)  = 3.9  meV and
$\Delta_L$(0) =  7.5 meV) in  some of their  local tunneling
spectra.  The same  inhomogeneity argument  could of  course
also explain such a superposition but a much more attractive
scenario  would be  a two-gap  model. Such  a model has been
first developed by Suhl {\it  et al.}\cite{suhl} in the case
of   overlapping  $s$-   an  $d$-   bands  in   conventional
superconductors (such  as V, Nb,  Ta). Experimental evidence
for the existence of two band superconductivity was obtained
by    tunneling   spectroscopy    in   Nb-doped    SrTiO$_3$
\cite{binning}. A  similar model has  been recently proposed
by Liu  {\it et al.}  for MgB$_2$. It  is here based  on the
coexistence  of a  two-dimensional Fermi  surface ($p_{x-y}$
orbitals)  perpendicular to  the $z$  direction and  a three
dimensional ($p_z$ bonding and  antibonding bands) one. This
model  is then  predicting  the  existence of  two different
energy gaps being respectively smaller  (for the 3D gap) and
slightly  larger (for  the 2D  gap) than  the expected  weak
coupling value.

Indications for  the existence of  two gaps in  MgB$_2$ have
also    been   found    in   specific    heat   measurements
\cite{bouquet,junod}  and  low  temperature Raman-scattering
\cite{chen} experiments  reveal two peaks  for two different
gaps. However, it is still  necessary to show that these two
gaps coexist up to $T_c$ in order to validate this scenario.
We  show   here  a  direct  and   clear  evidence  for  this
coexistence using  point-contact spectroscopy in  a magnetic
field.

Blonder, Tinkham and Klapwijk  (BTK) have developed a theory
\cite{blonder}   describing  the   electrical  transport  in
ballistic    contacts    between    a    normal    (N)   and
a superconducting  (S)  electrode  with  different  possible
interfaces  between them:  from a  pure conducting interface
where the Andreev reflection dominates  up to the well known
insulating  barrier (i.e.  the Giaever  tunneling case). The
most  important  consequence  of  this  theory  is  that any
point-contact  geometry will  be  able  to provide  a direct
spectroscopic  information about  the superconducting energy
gap. In  the pure Andreev  limit, if a  quasi-particle
is accelerated  by  an  applied  voltage  $V$  such that $|V| <
\Delta/e$,   a  direct   transfer  to   the  superconducting
electrode is forbidden and a hole is then retro-reflected in
the  normal electrode  in order  to allow  the formation  of
a Cooper  pair in  the superconductor.  The overall  current
(and differential  conductance $\sigma = dI/dV$)  for $|V| <
\Delta/e$  is then  twice higher  than the  value for $|V| >
\Delta/e$. In  the intermediate case  a dip appears  for $V$
= 0. Two  peaks are then  visible at $V  \sim \pm \Delta/e$.
The evolution  of the $dI/dV$  vs. $V$ curves  for different
interfaces  characterised with  the barrier  strength $Z$ is
schematically presented  in Fig. 1a. Note  that, as shown by
Deutscher   \cite{deutscher},  the   Giaever-like  tunneling
spectroscopy  will  provide  the  single particle excitation
energy (i.e.  the energy required to  break the Cooper pairs
required   for   superconductivity)   whereas   the  Andreev
reflection is associated with  the energy range of coherence
in  the  superconducting  state.  If  both  methods yield to
similar values  in conventional superconductors  they led to
very different energy gaps  in
underdoped high-$T_c$ oxides.
As the point-contact geometry  directly probes the coherence
of  the  superconducting  state,  it  is  probably  the most
adapted  technique to  determine the  superconducting energy
gap. Another advantage of  the point-contact spectroscopy is
that the normal electrode is pushed into the sample in order
to probe a clean surface.

    Point-contact   measurements  have   been  performed  on
polycrystalline  MgB$_2$ samples  with  $T_c$  = 39.3  K and
$\Delta T_c$  = 0.6 K.  A special point-contact  approaching
system with  a negligible thermal  expansion allows for  the
temperature and magnetic field measurements up to 100 K. The
point  contacts were  stable enough  to be  measured in  the
magnetic  field  of  a   superconducting  coil.  A  standard
lock-in  technique  at  400  Hz  was  used  to  measure  the
differential resistance as a  function of applied voltage on
the point contacts. The  microconstrictions were prepared by
pressing a copper tip (formed by electrochemical etching) on
the freshly polished surface  of the superconductor. MgB$_2$
samples  were prepared  from boron  powder (99,5  $\%$ pure,
Ventron) and magnesium  powder (98$\%$ Mg + 2  $\%$ KCl, MCP
Techn.), in relative proportion 1.05 :  2. A 2 g mass of the
mixed  powders was  introduced  into  a tantalum  tube, then
sealed by  arc melting under argon  atmosphere (purity 5N5).
The tantalum ampoule was  heated by high frequency induction
at 950  $^\circ$C for about  3 hours. After  cooling down to
room   temperature,  the   sample  was   analysed  by  X-ray
diffraction  and  Scanning  Electron  Microscope.  Among the
brittle dark grey MgB$_2$ powder (grain size $<$ 20 $\mu$m),
a few hard but larger grains (0.1  to 1 mm) were found. Laue
patterns show  evidence for only a  limited number of single
crystals   in    each   grain.   Resistivity    and   $a.c.$
susceptibility  measurements of  these larger  grains reveal
a particularly  abrupt  superconducting  transition ($\Delta
T_c \leq $0.6 K) indicating their high quality in comparison
with that of the fine powder.

Figure 1b  shows typical examples of  the conductance versus
voltage   spectra  obtained   for  the   various  Cu-MgB$_2$
junctions with  different barrier transparencies.  All shown
point-contact conductances have been normalized to the value
at  the   high-voltage  bias.  The   spectrum  had  a   more
tunneling-like  character  when  the  tip  first touched the
surface  (i.e.  with  a  barrier  resistance  $R  \sim  100$
$\Omega$) and then continuously transformed into a form with
a direct conductance  as the tip was  pushed into the sample
(down  to $R  \sim 6$  $\Omega$). Almost  all curves  reveal
a two-gap  structure  where  the   smaller  gap  maxima  are
displayed at around 2.8 mV and the large gap maxima at about
7 mV, placed symmetricaly around the  zero bias. Even, if in
some case only the smaller gap  is apparent (as shown in the
lowest curve of Fig. 1b), its width is hiding a contribution
of the second gap. Then, as  we show below, a magnetic field
can  suppress  the  smaller  gap  and  the  large  one  will
definitely emerge. All our curves could be fitted by the sum
of  the  two  BTK  conductances  $\alpha \sigma_S+(1-\alpha)
\sigma_L$  with  the  weight  factor  $\alpha$  varying from
$\sim 10\%$ to $\sim 90\%$  depending on the position of the
tip  (this scattering  of the  $\alpha $  value is  probably
related  to different  crystallographic orientations  at the
different  microconstrictions). We  thus definitely observed
two distributions with the  smaller gap scattered around 2.8
meV and the second one around 7 meV.

As  pointed out  above, the  smaller gap  could be caused by
a weakening of the  superconducting state possibly resulting
from a proximity  effect. That is why it is
important to show that it  still exists at high temperatures
in order  to validate the  two-gap scenario with  one single
critical  temperature. The  temperature dependencies  of the
point-contact spectra  have been examined  on five different
samples in about 10 spectra  with a clear two-gap structure.
Due to thermal smearing, the  two well resolved peaks in the
spectrum  merge  together   as  the  temperature  increases.
Consequently, the presence of two  gaps is not so evident in
the  raw  data  (see  Fig.  2a).  For  instance  at 25 K the
spectrum is  reduced to one smeared  maximum around the zero
bias.  Such a  spectrum itself  could be  fitted by  the BTK
formula with only one gap, but  the transparency
coefficient  $Z$  would have to change significantly in comparison
with the  lower temperatures and  moreover a large  smearing
factor $\Gamma$ would have been introduced.
However our data could be well
fitted  at   all  temperatures  by   the  sum  of   two  BTK
contributions with the transparency  coefficient $Z$ and the
weight  factor  $\alpha$  kept   constant  and  without  any
parallel leakage conductance. The point-contact conductances
of one spectrum at different  temperatures are shown in Fig.
2a together  with the corresponding BTK  fits. The resulting
energy  gaps $\Delta_L$  and $\Delta_S$  together with those
obtained for two other  point contacts with different weight
values $\alpha$  are shown in  Fig. 2b. The  error bars were
obtained   independently  for   each  spectrum   at a particular
temperature.

 Since it is evident from Fig. 2b that both gaps are closing
near  the same  bulk transition  temperature, our  data give
experimental support for the two gap model. In the classical
BCS theory,  an energy gap  with $\Delta_S$ =  2.8 meV could
not exist for a system  with $T_c$ above $2\Delta/3.5k \sim$
19 K. Moreover,  we obtained a very weakly  coupled gap with
$2\Delta_S/kT_c\simeq $ 1.7 and  a strongly coupled gap with
$2\Delta_L/kT_c \simeq$ 4.1 in  very good agreement with the
predictions   of  Liu   {\it  et   al.}  (a   3D  gap  ratio
$2\Delta_S/kT_c   \simeq  $   1.3   and   a  2D   gap  ratio
$2\Delta_L/kT_c  \simeq$ 4.0).  The temperature dependencies
are  in a  good agreement   with the  prediction of  the BCS
theory.  Small   deviations  from  this   theory  have  been
predicted  by  Liu  {\it  et  al.}  but these deviations are
within our error bars for  the large gap $\Delta_L$ while in
the case of the small gap $\Delta_S$ there is a tendency for
more rapid closing at higher temperatures near to $T_c$ (see
Fig. 2b) as expected theoretically.

Even much stronger evidence for the inherent presence of two
gaps in the superconductivity of  MgB$_2$ is obtained by our
magnetic field measurements. Figure 3 displays the effect of
the magnetic field up to 1 Tesla on the two gaps measured at
four different temperatures. At $T$ = 4.2 K the two gaps are
clearly  visible for  zero magnetic  field $H$  = 0  but the
peaks corresponding to the small gap are rapidly affected by
a magnetic field. The contribution  of this small gap almost
completely  disappears at  1 T  whereas the  large gap still
remains clearly visible. The other sets of spectra have been
recorded at  10, 20 and 30  K. As already shown  in Fig. 2a,
the spectra are so smeared out  above 20 K that the presence
of the two  gaps is not clearly visible and  we have to rely
on the fitting procedure. On the contrary, nothing like this
holds  when  we  apply   the  magnetic  field.  Indeed,  the
smaller-gap   contribution  to   the  overall  point-contact
conductance is  rapidly suppressed by the  field. As clearly
visible at  4.2 K and  10 K, this  suppression leads to  the
formation of a deep broad minimum in the conductance at zero
bias with increasing magnetic field  with some traces of the
smaller gap still surviving. Similarly, at 20 K and
30 K  the peaks corresponding  to the large  gap become much
better resolved  in a magnetic field.  This could not happen
if  there was  not a  contribution of  the small  gap in the
zero-field  point-contact  conductance   at  these  elevated
temperatures.  Indeed, a  one-gap spectrum  would only smear
out with increasing magnetic  field due to the pair-breaking
effect. This  effect thus unambiguously  shows that the  two
gaps coexist near $T_c$, thus  supporting the two band model
of superconductivity in MgB$_2$.

A quantitative analysis of the  magnetic field effect on the
two  gaps  is  a  problem  {\it  sui  generis}  and  will be
a subject of  our further study.  We suppose that  due to the
interaction  between the  3D and  2D bands  in MgB$_2$ there
will be not only one common  $T_c$ but also one $H_{c2}$ for
a particular  field  orientation.  Then,  the  rapid
suppression  of  the   smaller-gap  contribution at low fields
should  be
followed by  its weak presence  up to a common $H_{c2}$. If
single   crystals  are  available  it   would  be  very
interesting to get information  on the angular dependence of
the  gap structure  as a  function of  magnetic field.  This
could  shed  some  light  on  the very  important  but so far
contradictory   problem  of   the  anisotropy   of  $H_{c2}$
\cite{lima}  and its  relation to  the gaps.  On the spectra
obtained so far we never  observed a full suppression of the
gap structure below 10 Tesla at 4.2 K.

The specific heat measurements  performed by Bouquet {\it et
al.}  \cite{bouquet}  revealed  a  strong  reduction  of the
electronic density of states  at low temperatures suggesting
the presence of a second small energy gap which, in striking
similarity with our direct observations, could be suppressed
by a $\sim $ 1 Tesla magnetic field. Specific heat is a bulk
thermodynamic  quantity  and  the  excellent  consistency of
these  measurements with  our spectroscopic  results further
demonstrates  that the  existence of  these two  gaps is  an
intrinsic property  of MgB$_2$. The presence of  two gaps
has  been recently detected in the specific heat data for three
differently prepared samples  \cite{bouquet2}. Recently also
a body  of  optical  and   microwave  measurements  is
appearing (e.g.  Tu {\it et  al.} \cite{tu}) on MgB$_2$
samples  of  different  forms  showing that the
smaller gap  is a bulk property.  Indirect observations of
two gaps were also  obtained from the temperature dependence
of the specific heat in the conventional superconductors Nb,
Ta  and V  \cite{shen} but   to the  best of  our knowledge,
a clear and direct observation for  the presence of two gaps
existing up to  the same $T_c$ was never  observed before by
spectroscopic measurements.  The possibility that  these two
gaps  have  different  dimensionalities  makes  this  system
particularly attractive for further studies.

In  conclusion,  we  have  obtained  a  strong  experimental
evidence for the  existence of two gaps closing  at the same
bulk  $T_c$  of  the  MgB$_2$  superconductor.  The  regular
observation of  this effect in  our spectra and  the support
for  it  by  other  techniques  probing  the  the  sample on
a different scale indicate that this is an inherent property
of the material.

\acknowledgments
This work has  been supported by the Slovak  VEGA grant No.1148.
We would like to thank D. Roditchev for useful discussions.


\newpage

{\bf FIGURE CAPTIONS}

\vspace{1cm}

FIG.  1.  a)  Numerical  simulation  of  the  BTK  model  at
different values  of the barrier  strength $Z$, representing
behavior of  the point-contact spectra for  $\Delta =$ 7 meV
between  Giaver  tunneling  ($Z  =$  10)  and  clean Andreev
reflexion  ($Z  =  $  0)  at  $T=$  4.2 K. b) Experimentally
observed evolution of the Cu - MgB$_2$ point-contact spectra
at $T = 4.2$ K (full lines). The upper curves are vertically
shifted  for  the  clarity.  Dotted  lines  display  fitting
results for  the thermally smeared BTK  model with $\Delta_S
= 2.8  \pm 0.1$  meV, $\Delta_L  =  6.8  \pm 0.3  $ meV  for
different barrier transparencies and weight factors.

\vspace{1cm}

FIG.  2.  a)  Differential  conductances  of  Cu  -  MgB$_2$
point-contact  measured  (full  lines)  and  fitted  (dotted
lines)  for the  thermally  smeared  BTK model  at indicated
temperatures. The  fitting parameters $\alpha =  0.71$, $Z =
0.52 \pm 0.02$  had the same values at  all temperatures. b)
Temperature dependencies of  both energy gaps ($\Delta_S(T)$
- solid  symbols, $\Delta_L(T)$  - open  symbols) determined
from  the  fitting  on  three  different  point-contacts are
displayed   with  three   corresponding  different  symbols.
$\Delta_S(T)$ and  $\Delta_L(T)$ points determined  from the
same  contact are  plotted with  the same  (open and  solid)
symbols. Full lines represent BCS predictions.

\vspace{1cm}

FIG.  3. Experimentally  observed influence  of the  applied
magnetic field  on the two  gap structure of  the normalized
point-contact  spectra  at   indicated  temperatures.  These
spectra clearly  reveal that both gaps  exists in zero field
up to $T_c$  as shown by the rapid  suppression of the small
gap  structure ($\Delta_S  = 2.8$  meV) with  magnetic field
which leads to a broad deep minimum at zero bias.

\newpage
\onecolumn

\begin{figure}
\epsfverbosetrue
\epsfxsize=10cm
\epsfysize=16cm
\begin{center}
\hspace{550mm}
\epsffile{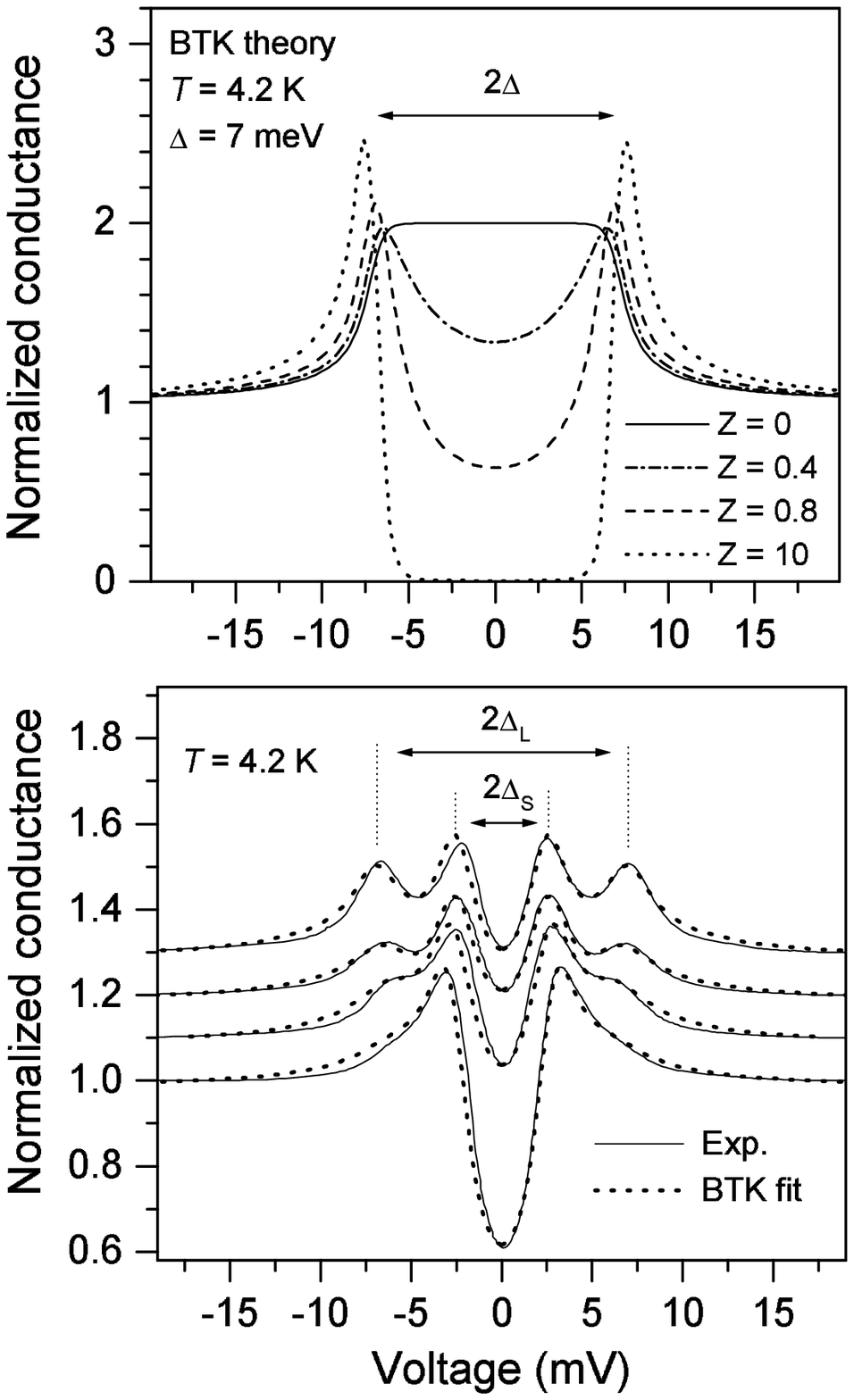}
\vspace{2mm}
\end{center}
\caption{
a)  Numerical  simulation  of  the  BTK  model  at
different values  of the barrier  strength $Z$, representing
behavior of  the point-contact spectra for  $\Delta =$ 7 meV
between  Giaver  tunneling  ($Z  =$  10)  and  clean Andreev
reflexion  ($Z  =  $  0)  at  $T=$  4.2 K. b) Experimentally
observed evolution of the Cu - MgB$_2$ point-contact spectra
at $T = 4.2$ K (full lines). The upper curves are vertically
shifted  for  the  clarity.  Dotted  lines  display  fitting
results for  the thermally smeared BTK  model with $\Delta_S
= 2.8  \pm 0.1$  meV, $\Delta_L  =  6.8  \pm 0.3  $ meV  for
different barrier transparencies and weight factors.
 }
\end{figure}

\newpage

\begin{figure}
\epsfverbosetrue
\epsfxsize=10cm
\epsfysize=16cm
\begin{center}
\hspace{550mm}
\epsffile{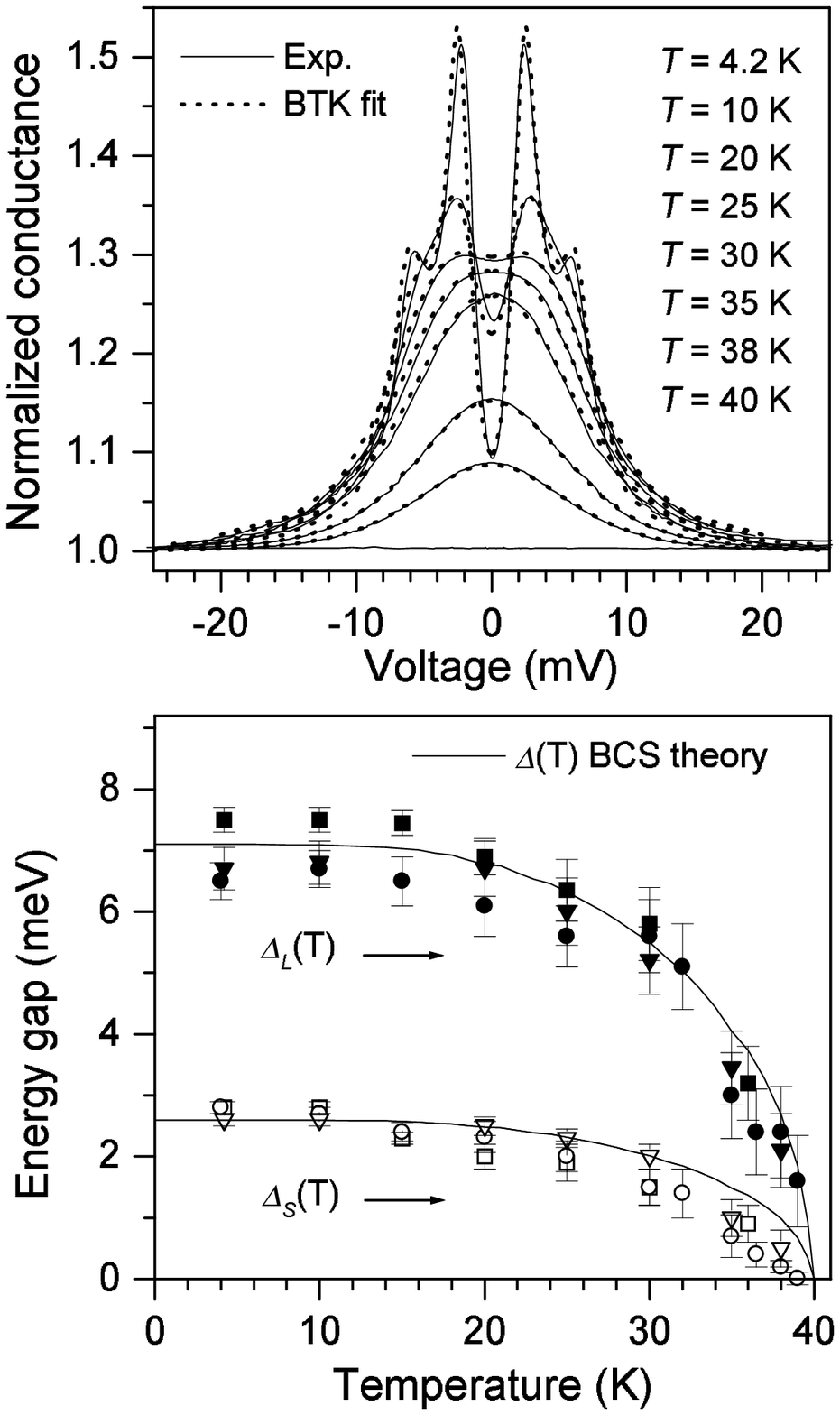}
\vspace{2mm}
\end{center}
\caption{ a)  Differential  conductances  of  Cu  -  MgB$_2$
point-contact  measured  (full  lines)  and  fitted  (dotted
lines)  for the  thermally  smeared  BTK model  at indicated
temperatures. The  fitting parameters $\alpha =  0.71$, $Z =
0.52 \pm 0.02$  had the same values at  all temperatures. b)
Temperature dependencies of  both energy gaps ($\Delta_S(T)$
- solid  symbols, $\Delta_L(T)$  - open  symbols) determined
from  the  fitting  on  three  different  point-contacts are
displayed   with  three   corresponding  different  symbols.
$\Delta_S(T)$ and  $\Delta_L(T)$ points determined  from the
same  contact are  plotted with  the same  (open and  solid)
symbols. Full lines represent BCS predictions.
}
\end{figure}

\begin{figure}
\epsfverbosetrue
\epsfxsize=12cm
\epsfysize=13cm
\begin{center}
\hspace{550mm}
\epsffile{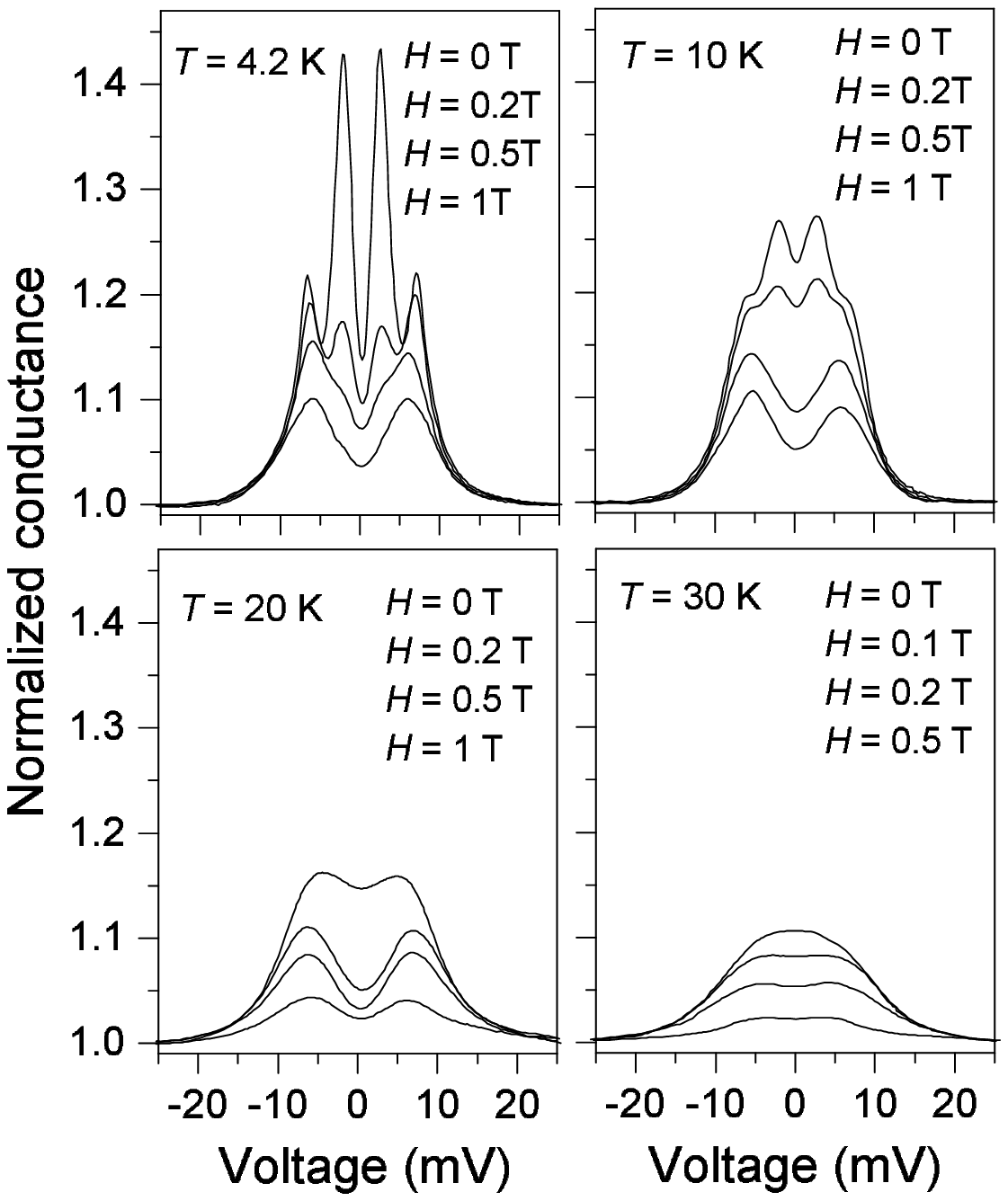}
\vspace{2mm}
\end{center}
\vspace{2mm}
\caption{Experimentally  observed influence  of the  applied
magnetic field  on the two  gap structure of  the normalized
point-contact  spectra  at   indicated  temperatures.  These
spectra clearly  reveal that both gaps  exists in zero field
up to $T_c$  as shown by the rapid  suppression of the small
gap  structure ($\Delta_S  = 2.8$  meV) with  magnetic field
which leads to a broad deep minimum at zero bias.}
\end{figure}

\end{document}